\documentclass[fleqn,usenatbib]{mnras}

\usepackage{newtxtext,newtxmath}

\usepackage[T1]{fontenc}

\DeclareRobustCommand{\VAN}[3]{#2}
\let\VANthebibliography\thebibliography
\def\thebibliography{\DeclareRobustCommand{\VAN}[3]{##3}\VANthebibliography}

\usepackage{graphicx}	
\usepackage{amsmath}	
\usepackage{tabularx}

\title{Interstellar scintillation of sources B0821+394 and B1812+412 as observed by the LPA LPI radio telescope}

\author[S. A. Tyul'bashev et al.]{
S. A. Tyul'bashev $^{1}$\thanks{E-mail: serg@prao.ru}
I.V. Chashei,$^{1}$
I.A. Grishanova,$^{2}$
G.E. Tyul'basheva,$^{3}$
I.A. Subaev,$^{1}$
\\
$^{1}$ Pushchino Radio Astronomy Observatory, Astro Space Center, Lebedev Physical Institute, Russian Academy of Sciences, Pushchino, 142290 Russia \\
$^{2}$ Pushchino brunch of Russian Biotechnological University (Rosbiotech), Pushchino, Moscow reg., Russia \\
$^{3}$ Institute of Mathematical Problems of Biology RAS (IMPB RAS) brunch of Keldysh Institute of Applied Mathematics of Russian Academy of Sciences, \\
Pushchino, Moscow reg., Russia \\
}

\date{ }

\pubyear{ }

\begin{document}
\label{firstpage}
\pagerange{\pageref{firstpage}--\pageref{lastpage}}
\maketitle

\begin{abstract}
The search for long-term variability of compact components of radio sources B0821+394 and B1812+412 over an interval of 10 years was carried out. The LPA LPI radio telescope with an operating frequency of 111 MHz was used for observations. According to our estimates, the characteristic time of variability for both sources is 1.5-2.5 years. It is shown that the observed variability is not related to intrinsic variations in the radiation flux, but is due to refractive scintillation on inhomogeneities of the interstellar medium. From the obtained upper estimates of the apparent angular dimensions of the sources, it follows that the main contribution to the scattering of radio emission is made by turbulent plasma concentrated in sufficiently thin screens, the distance to which does not exceed 300-400 pc.
\end{abstract}




\section{Introduction}

For the first time, the low-frequency variability observed in compact radio sources was reported by \citeauthor{Hunstead1972} (\citeyear{Hunstead1972}). In the observations, which took place at a frequency of 408 MHz (wavelength 74 cm), he found a variability, the nature of which was not clear. In Hunsted's work, it was noted that the behavior of sources (variability of flux density) at low frequencies differs from their behavior at high frequencies.

Following Hunsted, a number of authors \citeauthor{Cotton1976} (\citeyear{Cotton1976}); \citeauthor{Fanti1981} (\citeyear{Fanti1981}); \citeauthor{Slee1988} (\citeyear{Slee1988}); \citeauthor{McGilchrist1990} (\citeyear{McGilchrist1990}) also searched for low-frequency (80-408 MHz) variability. In the work \citeauthor{Cotton1976} (\citeyear{Cotton1976}) it was shown that strong variability at a frequency of $\sim$~372 MHz in a complete stream sample ($\sim~$1500 sources) is very rare - one source out of three hundreds (0.33\%), and variability in general is shown by 1.5\% of sources in the selection. A similar result was obtained in \cite{McGilchrist1990} with observations at 151 MHz. In this work, a search was carried out for variability in sources having dimensions of the order of an angular minute, and variability was found in 1.1\% of sources. For a sample of deliberately compact sources selected at 408 MHz (\citeauthor{Fanti1981}, \citeyear{Fanti1981}), the proportion of sources with variability increased to 25\%.

The observed low-frequency variability raised the question of its nature. If the causes of the variability are internal, the sources should contain a compact component (the active nucleus of the galaxy; AGN), the flux density of which varies on time scales of months or years. That is, this component should be visible not only at low (meter range), but also at high (decimeter and centimeter range) frequencies. The standard model of an adiabatically expanding source (\citeauthor{Shklovsky1965}, \citeyear{Shklovsky1965}; \citeauthor{Laan1966}, \citeyear{Laan1966}; \citeauthor{Marscher1985}, \citeyear{Marscher1985}; \citeauthor{Mitchell1994}, \citeyear{Mitchell1994}) shows that the lower the frequency of observations, the smaller the proportion of variable flow should be. At the same time, there should be a correlation of the alternating flux observed at high and low frequencies. However, a comparison of the correlated variability of sources often shows that there is no relationship between variability at high and low frequencies.

The observed low-frequency variability of the AGN may be related to refractive scintillating in the interstellar medium (\citeauthor{Shapirovskaya1978}, \citeyear{Shapirovskaya1978}; \citeauthor{Rickett1986}, \citeyear{Rickett1986}). In this case, there will be no correlation of low-frequency and high-frequency variability, but for the appearance of interstellar scintillation, the sources must have components of small angular dimensions. A mixture of internal (intrinsic) and external (due to the interstellar medium) causes of variability is also very likely. Thus, identifying the nature of the variability observed at low ($< 400$~MHz) frequencies is a job that requires long-term series of observations, accurate measurements of the flux density and analysis of observations made at different frequencies.

In 2019, an attempt was made to search for variability at a frequency of 111 MHz (wavelength 2.7 meters) for three sources with flat spectra (\citeauthor{Tyul'bashev2019}, \citeyear{Tyul'bashev2019}), previously observed by the method of interplanetary scintillating in a sample of 48 sources (\citeauthor{Tyul'bashev2005}, \citeyear{Tyul'bashev2005}). The light curves (dependence of the flux density on time) obtained over a 5-year interval showed probable variability for all sources. However, the characteristic scale of variability ($\sim$2-3 years), comparable to the duration of a number of observations, did not allow for data analysis. In this paper, the analysis of the light curves obtained over an interval of 10 years is carried out.

\section{Observations and processing}

The observations were carried out on the Large Phased Array (LPA) of the Lebedev Physical Institute (LPI). The LPA LPI is a meridian instrument having two independent directional patterns. The LPA1 radio telescope is designed for observations of individual radio sources with high time-frequency resolution. The LPA3 diagram has 128 uncontrolled (stationary) beams having fixed directions in the meridian plane and covering declinations from $-9^{\circ}$ to $+55^{\circ}$. The central frequency of observations is 111 MHz, the received frequency band is 2.5 MHz, the sampling is 0.1 s. Since the middle of 2012, monitoring observations of the celestial sphere on LPA3 have been carried out daily and around the clock on a 6-channel receiver with a channel width of 415 kHz. The antenna pattern size is approximately $0.5^{\circ}\times 1^{\circ}$. The transit time of the source at half power takes about 3.5 minutes per day (for one observation session). Instant sky coverage - 50 sq.deg. (in 128 beams), the sky coverage per day of observations is 17000 sq.deg.

To control the quality of observations, the antenna is switched off 6 times a day, and a calibration signal (calibration step) of a known temperature is supplied to the antenna-feeder system using a noise generator. Calibration steps allow you to align signals in frequency channels both at intervals of 4 hours (between two steps) and between steps for the entire observation period if you take one of the steps as a reference. Since the temperature of the step does not change with time, it is possible to align the signals at intervals of years.

Data for searching for variability from extragalactic radio sources were taken from observations on LPA3. The detailed methodology for processing observations is described in \citeauthor{Tyul'bashev2019} (\citeyear{Tyul'bashev2019}). Here we give the basic steps:

- we cut out the source from the monitoring data, and remove the interference. If the recording quality remains low after removing the interference, we exclude the recording from further processing;

- we use a calibration step to equalize the gain;

- we estimate the intensity of the calibration and investigated sources in temperature units;

- in the resulting light curve, we remove short-duration interference using a three-point median filter;
 
 - we average the intensities over a six-month interval (before this step, the studied and calibration sources are treated the same way);
 
 - assuming that the flux density of the calibration sources does not change, and the visible deviations in their light curves are related to the physical state of the antenna, we divide the relative intensities of the studied sources by the relative intensities of the calibration sources and obtain the final light curve;
 
 - using the known values of the flux density of calibration sources, we translate the intensities on the light curves from conventional intensity units to Jy ones.

\begin{figure*}
\begin{center}
	\includegraphics[width=0.8\textwidth]{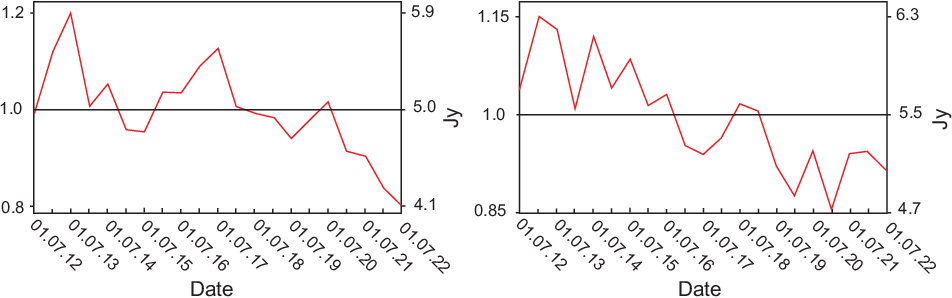}
    \caption{The light curves of the sources B0821+394 and B1812+412. The left vertical axis on the left and right panels shows the intensity in fractions of one (the average intensity value over the entire light curve is taken as one), the right vertical axis shows the integral density of the source flux in Jy. The time is marked on the lower horizontal axis in the format day, month, year. The time step between the labels on the horizontal axis is six months.}
    \label{fig:fig1}
\end{center}
\end{figure*}

Using the scheme of operation described above, we obtained light curves over a 10-year interval for sources B0821+394 (J0824+3916; 4C+39.23) and B1812+412 (J1814+4113; 4C+41.37). These sources have flat spectra at high frequencies and were previously studied at the LPA by the method of interplanetary scintillating in a sample of sources with flat spectra (see (\citeauthor{Tyul'bashev2005}, \citeyear{Tyul'bashev2005}) and references therein). All sources are also characterized by the fact that, in addition to the compact component, their integral flux density is determined on the LPA, which makes it possible to determine the proportion of energy in the compact component at known angular dimensions. The obtained light curves reflecting the integral density of the sources at semi-annual intervals are shown in Fig.~\ref{fig:fig1}.

\section{Analysis of the results}

\subsection{Time scales of variability}

The light curves of the sources B0821+394 and B1812+412 show intensity differences of $\sim$15-20\% of the average value. According to the work of \citeauthor{Tyul'bashev2019} (\citeyear{Tyul'bashev2019}), the expected accuracy of flux density estimates when averaging intensity estimates over six months reaches one percent, therefore all observed intensity differences should be determined either by internal (intrinsic) variability or external (interstellar scintillating) variability. It can be seen from Fig.~\ref{fig:fig1} that the characteristic scales of intensity variability are 1.5-2.5 years. Integral flux densities $S_{int} = 5$~Jy and 5.5 Jy, minimum and maximum values of flux density on the light curve are 4.1-5.9 Jy and 4.7-6.3 Jy for B0821+394 and B1812+412, respectively.

\subsection{Temperature and angular dimensions of sources in case of intrinsic variability}

If the observed variability is related to changes in the source itself, its temperature and angular dimensions can be estimated. The angular dimensions are determined from the linear dimensions of the sources (the upper estimate of the size is related to the characteristic scale of variability) and the distance to the source (see, for example, the NED database - https://ned.ipac.caltech.edu/). The flux densities of compact components are determined in \citeauthor{Tyul'bashev2019} (\citeyear{Tyul'bashev2019}). According to the formula given in \citeauthor{Kovalev2005} (\citeyear{Kovalev2005}), the brightness temperature is defined as:

\begin{equation}
	T_b = 5.44 \times 10^9 \times \left[ \frac{S_c (1+z)}{\theta^2} \right] \, K,	
	\label{eq:1}		
\end{equation}

where $S_c$ is the flux density of a compact radio source in Jy, $\theta$ is the source size in milliarkseconds (ms). Based on the flux density of the compact component $S_c = 0.5-1.0$~Jy \cite{Tyul'bashev2019}, upper estimates of the angular dimensions of the components $\theta = 0.02$ and 0.015 ms (calculated using NED), redshifts $z = 1.215$ and 1.564 for B0821+394 and B1812+412, respectively, we get $T_b > 10^{13}$~K. This temperature is an order of magnitude higher than the Compton limit of $10^{12}$~K, so the intrinsic variability of the sources seems unlikely.

\subsection{Source parameters under the assumption of refractive scintillation}

Radiation from the components of an extragalactic radio source passes through the interstellar medium in the halo and in the plane of the Galaxy, and is scattered along the way. Thus, the interstellar medium determines the minimum visible angular dimensions of the sources. According to \citeauthor{Rickett1986} (\citeyear{Rickett1986}), these minimum dimensions (scattering angle) can be defined as

\begin{equation}
	\Theta \approx 8 \lambda^2 (cosec \, b)^{0.5} \, 	
		\label{eq:2}		
	\end{equation}

where $\lambda$ and $b$ are the wavelength in meters and the galactic latitude at which the source is visible. In our case, $\lambda = 2.7$~m, and the galactic latitudes for B0821+394 and B1812+412 are, respectively, $34.714^{\circ}$ and $23.841^{\circ}$. The estimates of the scattering angle are 77 ms and 92 ms (upper estimates of the observed source sizes). The minimum scattering for extragalactic radio sources should be observed in directions with a minimum of the interstellar medium. Thus, the minimum scattering angle at a frequency of 111 MHz, according to formula 2, should be about 58 ms. According to measurements of the angular dimensions of pulsars and quasars carried out at a frequency of 102.5 MHz (\citeauthor{Artyukh1989}, \citeyear{Artyukh1989}), the minimum recorded angular dimensions were 60 ms. Thus, the values of the scattering angles obtained by formula (2) and the experimental results obtained for the same frequencies on the LPA LPI radio telescope converge. The apparent angular dimensions of both sources, due to the inevitable scattering in the interstellar medium, will be hundreds of times larger than those estimated from the characteristic scales of variability. It is interstellar scattering that will limit the maximum angular resolution of extragalactic radio sources in the meter wavelength range.

Let's estimate at what angular dimensions the interstellar scintillation mode will be weak, and the scintillation will not be saturated. According to formula 2.3 in \citeauthor{Rickett1990} (\citeyear{Rickett1990}):

\begin{equation}
	\theta_{weak} \approx 8\times 10^{-6} (L_{kpc} f_{GHz})^{-1/2} \, arcsec,		
		\label{eq:3}		
	\end{equation}

where $\theta_{weak}$ is the angular size of the source, which distinguishes the modes of weak and saturated scintillating, $L_{kpc}$ and $f_{GHz}$ is the distance to the screen (the value of 5 kpc was taken for estimates) and the frequency of observations (0.111 GHz). The angular size of $\theta_{weak} \approx 0.01$~ms, which is many times smaller than the scattering disk. Therefore, if the observed variability is associated with interstellar scintillating in saturation mode, it will be refractive scintillating.

It is also possible to estimate the proportion of energy emitted by a compact component. It is determined by the visible scintillation index of the sources:

\begin{equation}
		m = \frac{\langle (I - \langle I \rangle )^2 \rangle}{ \langle I \rangle},		
		\label{eq:4}		
	\end{equation}

where $I$ is the intensity in conventional units. The scintillation indices of $m = 0.26$ (B0821+394) and $m = 0.29$ (B1812+412) are approximately equal. Since the sources are compact, the theoretical value of the scintillation index $m_0$, attributed to the compact component, will be $m_0 = 1$, and in this case the apparent value of m directly determines the proportion of energy of the scintillating component. Therefore, the energy in the compact (scintillating on interstellar plasma) component $S_c = 1.3$~Jy and 1.6 Jy for B0821+394 and B1812+412.

Based on the fact that the observed variability is associated with refractive scintillating, it is possible to estimate the apparent angular dimensions of the sources (\citeauthor{Mantovani1990}, \citeyear{Mantovani1990}):

\begin{equation}
	\frac{L_0}{\nu} = 1.7 \tau \frac{\sin b}{\theta_m},			
		\label{eq:5}		
	\end{equation}

where $L_0$ is half the thickness of the translucent layer in pc towards the Galactic Pole, $\nu$ is the speed in km/s, $\tau$ is the characteristic time of variability in days, $\theta_m$ is the measured angular size in ms. Let $L_0 = 5000$~pc, that is, we consider the case of a thick statistically homogeneous medium when radiation passes through the halo and part of the galactic plane), $\nu = 50$~km/s, $\tau = 730$~days (2 years $= 365\times 2 = 730$). Then we get an estimate of the angular size of $\theta_m\approx 6$~ms for both sources. Since the scattering angle associated with the interstellar medium and estimated by formula (2) is an order of magnitude larger (see above), the observed variability cannot be associated with a thick medium.

The natural explanation for the difference in the estimates of the scattering angle is that the main modulation of radiation occurs in a thin screen located close enough to the Sun. The existence of such screens in the interstellar medium was shown, for example, in the works \citeauthor{Smirnova2014} (\citeyear{Smirnova2014}); \citeauthor{Shishov2017} (\citeyear{Shishov2017}) based on the analysis of pulsar measurement data. In the cases considered in this paper, if the distance to the screen is 500 pc, scattering by turbulence in the screen will lead to visible angular dimensions of about 60 mas. Based on estimates of the scattering angles, that is, the minimum possible visible angular dimensions, 77 ms and 92 ms (for B0821+394 and B1812+412), the distance to the screens will be 390 pc and 325 pc. The estimated distance to the screens is the upper limit, whereas real screens can be located even closer.

Scattering on the interstellar medium determines the apparent angular dimensions of the sources and at the same time limits the apparent temperatures. Taking into account the scattering, it is possible to estimate the actual temperature of the sources, and even to identify their structure in the presence of multi-frequency VLBI observations (see, for example, \citeauthor{Pilipenko2018} (\citeyear{Pilipenko2018}). However, to obtain temperature estimates based on observations at the LPA, estimates of the angular dimensions of sources from the VLBI observations conducted at a frequency of 111 MHz are required.

\section{Discussion of the results and conclusion}

Both sources have numerous VLBI observations. Source B0821+394 is included in the sample of sources observed in the MOJAVE project (https://www.cv.nrao.edu/MOJAVE/sample.html; (\citeauthor{Lister2013}, \citeyear{Lister2013})). The published light curve shows that at a frequency of 15 GHz, the flux density fluctuates near $S_c=1$~Jy, which is very close to the average value of $S_c = 1.3$~Jy given in the paragraph above over a 10-year interval. According to the analysis carried out in \citeauthor{Tyul'bashev2019} (\citeyear{Tyul'bashev2019}), the main flux density is determined by a component having an angular size of 1-2 ms, with a total source size of 5 ms at a frequency of 15 GHz (\citeauthor{Lister2013}, \citeyear{Lister2013}). That is, the spectrum of the compact component B0821+394 remains flat over a wide range of wavelengths. According to VLBI observations of B1812+412 at 5 GHz (\citeauthor{Henstock1995}, \citeyear{Henstock1995}), the main energy comes from a component with a size of 0.3$\times$0.4 ms with a total source length of 10 ms.

The dimensions of the compact components of the sources B0821+394 and B1812+412 are such that the scattering of these components on the interstellar medium should be observed both for the case of a thick medium and for the case of a thin screen. The details determining the main density of the source flux are resolved in the VLBI observations. Therefore, the observed intensity variations cannot be related to their own variability. Recall that for the case of intrinsic variability, the parts must be less than 0.02 ms.

Thus, refractive interstellar scintillating makes it possible to determine the angular size of the source if there is a thin screen in the direction of the source and the distance to the screen is known, or to estimate the distance to the screen if the observed variability can be associated with interstellar scintillating, and the apparent angular dimensions of the source are known. Coincidentally, the resolution of the methods of interstellar scintillating (in this paper) and interplanetary scintillating (\citeauthor{Artyukh1989}, \citeyear{Artyukh1989}) during observations at the LPA LPI are close. A thick screen can apparently be excluded from consideration for all or almost all directions in the sky. Interstellar scintillating is probably related to turbulence concentrated in fairly thin screens. Thin screens can be positioned in different directions at different distances, and these distances can be estimated by taking the apparent angular dimensions of the sources, determined by interplanetary scintillating or using VLBI methods.

In addition to several "waves of variability", Fig.~\ref{fig:fig1} shows a trend of decreasing flux density over time for both sources. The minimum characteristic time of variability for this trend is 10 years. If the observed drop in flux density is associated with refractive scintillating, then this may indicate another thin screen located at a distance not exceeding 100 pc.

In conclusion, we note the main results:

- observations over a 10-year interval made it possible to reliably determine the long-term variability of B0821+394 and B1812+412;
 
- the estimates of the scattering angles made above show that the observed variability with characteristic times of 1.5-2.5 years is associated with interstellar refractive scintillation on thin screens, the distance to which does not exceed 400 pc.

{ACKNOWLEDGMENTS}

The authors thank the LPA antenna group for the constant support of the radio telescope in monitoring mode and L.B. Potapova for help in preparing the paper.

\bibliographystyle{mnras}

\begin{thebibliography}{}
\bibitem[Artyukh \&	Smirnova(1989)]{Artyukh1989}
{Artyukh} V.~S.,  {Smirnova} T.~V.,  1989, Soviet Astronomy Letters, {15, 344}
\bibitem[Cotton(1976)]{Cotton1976}
{Cotton} W.~D.,  1976, \apjs,  {32, 467}
\bibitem[Fanti et~al.(1981)]{Fanti1981}
{Fanti} C.,  {Fanti} R.,  {Ficarra} A.,  {Montovani} F.,  {Padrielli} L.,
{Weiler} K.~W.,  1981, \aaps,  {45, 61}
\bibitem[Henstock et~al.(1995)]{Henstock1995}
{Henstock} D.~R.,  {Browne} I.~W.~A.,  {Wilkinson} P.~N.,  {Taylor} G.~B.,
{Vermeulen} R.~C.,  {Pearson} T.~J.,   {Readhead} A.~C.~S.,  1995, \apjs,  {100, 1}
\bibitem[Hunstead(1972)]{Hunstead1972}
{Hunstead} R.~W.,  1972, \aplett,  {12, 193}
\bibitem[Kovalev et~al.(2005)]{Kovalev2005}
{Kovalev} Y.~Y.,  et~al., 2005, \aj,  {130, 2473}
\bibitem[Laan(1966)]{Laan1966}
H. van der Laan, 1966, Nature (London, U.K.) {211, 1131} 
\bibitem[Lister	et~al.(2013)]{Lister2013}
{Lister} M.~L.,  et~al., 2013, \aj, {146, 120}
\bibitem[Mantovani et~al.(1990)]{Mantovani1990}
{Mantovani} F.,  {Fanti} R.,  {Gregorini} L.,  {Padrielli} L.,   {Spangler} S.,
1990, \aap, {233, 535}
\bibitem[Marscher \& Gear(1985)]{Marscher1985}
{Marscher} A.~P.,  {Gear} W.~K.,  1985, \apj,  {298, 114}
\bibitem[McGilchrist \& Riley(1990)]{McGilchrist1990}
{McGilchrist} M.~M.,  {Riley} J.~M.,  1990, \mnras,  {246, 123}
\bibitem[Mitchell et~al.(1994)]{Mitchell1994}
{Mitchell} K.~J.,  {Dennison} B.,  {Condon} J.~J.,  {Altschuler} D.~R.,
{Payne} H.~E.,  {O'Dell} S.~L.,   {Broderick} J.~J.,  1994, \apjs,  {93, 441}
\bibitem[Pilipenko et~al.(2018)]{Pilipenko2018}
{Pilipenko} S.~V.,  et~al., 2018, \mnras, {474, 3523}
\bibitem[Rickett(1986)]{Rickett1986}
{Rickett} B.~J.,  1986, \apj,  {307, 564}
\bibitem[Rickett(1990)]{Rickett1990}
{Rickett} B.~J.,  1990, \araa, {28, 561}
\bibitem[Shapirovskaya(1978)]{Shapirovskaya1978}
{Shapirovskaya} N.~Y.,  1978, \sovast,  {22, 544}
\bibitem[Shishov et~al.(2017)]{Shishov2017}
{Shishov} V.~I.,  {Smirnova} T.~V.,  {Gwinn} C.~R.,  {Andrianov} A.~S.,
{Popov} M.~V.,  {Rudnitskiy} A.~G.,   {Soglasnov} V.~A.,  2017, \mnras,  {468, 3709}
\bibitem[Shklovsky(1965)]{Shklovsky1965}
{Shklovsky} J.,  1965, \nat,  {206, 176}
\bibitem[Slee \& Siegman(1988)]{Slee1988}
{Slee} O.~B.,  {Siegman} B.~C.,  1988, \mnras, {235, 1313}
\bibitem[Smirnova et~al.(2014)]{Smirnova2014}
{Smirnova} T.~V.,  et~al., 2014, \apj, {786, 115}
\bibitem[Tyul'bashev \& Augusto(2005)]{Tyul'bashev2005}
{Tyul'Bashev} S.~A.,  {Augusto} P.,  2005, \aap,  {439, 963}
\bibitem[Tyul'bashev et~al.(2019)]{Tyul'bashev2019}
{Tyul'bashev} S.~A.,  {Golysheva} P.~Y.,  {Tyul'bashev} V.~S.,   {Subaev}
I.~A.,  2019, Astronomy Reports,  {63, 920}
	\end{thebibliography}

\end{document}